\documentclass[prl,aps,superscriptaddress,fleqn,showpacs,twocolumn]{revtex4}
\usepackage{amsmath,amssymb,graphicx}
\begin{document}

\title{Reconstruction of protein structures from a vectorial representation}

\author{Markus~Porto}
\affiliation{Max-Planck-Institut~f\"ur~Physik~komplexer~Systeme,
             N\"othnitzer~Stra{\ss}e~38, 01187~Dresden, Germany}
\affiliation{Institut~f\"ur~Theoretische~Physik,
             Technische~Universit\"at~Dresden, 01062~Dresden, Germany}

\author{Ugo~Bastolla}
\affiliation{Centro~de~Astrobiolog{\'\i}a~(INTA-CSIC),
             28850~Torrej\'on~de~Ardoz, Spain}

\author{H.~Eduardo~Roman}
\affiliation{Dipartimento~di~Fisica and INFN, Universit\`a~di~Milano,
             Via~Celoria~16, 20133~Milano, Italy}

\author{Michele~Vendruscolo}
\affiliation{Department~of~Chemistry, University~of~Cambridge,
             Lensfield~Road, Cambridge CB2~1EW, UK}

\date{September 5, 2003, revised January 7, 2004}

\begin{abstract}
We show that the contact map of the native structure of globular proteins can
be reconstructed starting from the sole knowledge of the contact map's
principal eigenvector, and present an exact algorithm for this purpose. Our
algorithm yields a unique contact map for all $221$ globular structures of
PDBselect25 of length $N \le 120$. We also show that the reconstructed contact
maps allow in turn for the accurate reconstruction of the three-dimensional
structure. These results indicate that the reduced vectorial representation
provided by the principal eigenvector of the contact map is equivalent to the
protein structure itself. This representation is expected to provide a useful
tool in bioinformatics algorithms for protein structure comparison and
alignment, as well as a promising intermediate step towards protein structure
prediction.
\end{abstract}


\maketitle

\noindent\textit{Introduction.} -- Despite several decades of intense research,
the reliable prediction of the native state of a protein from its sequence of
amino acids is still a formidable challenge \cite{koh03}. Every two years the
state of the art is assessed by the CASP experiment \cite{CASP}. In the most
detailed predictions, the list of the Cartesian coordinates of all the atoms of
the protein molecule are provided. Since the structure of a protein can be also
represented as a contact map \cite{lifson79}, lower-resolution predictions can
be limited to the determination of inter-residue contacts \cite{fariselli01}.

Ideally, one would like to predict the structure of a protein using the
representation that is encoded in the most straightforward way in the sequence
of amino acids. Due to the vectorial nature of the sequence, one may guess that
the simplest representation to predict should be vectorial as well. In this
Letter, we show that the principal eigenvector (PE) of the contact map (CM) of
the native structure is equivalent to the CM itself, and therefore provides a
very promising vectorial representation of protein structures. This vectorial
representation is also expected to improve bioinformatics algorithms for
protein structure alignment as well as alignment of protein sequence with a
database of structures (fold recognition).

\begin{figure}[t]
\centerline{\includegraphics[scale=0.52]{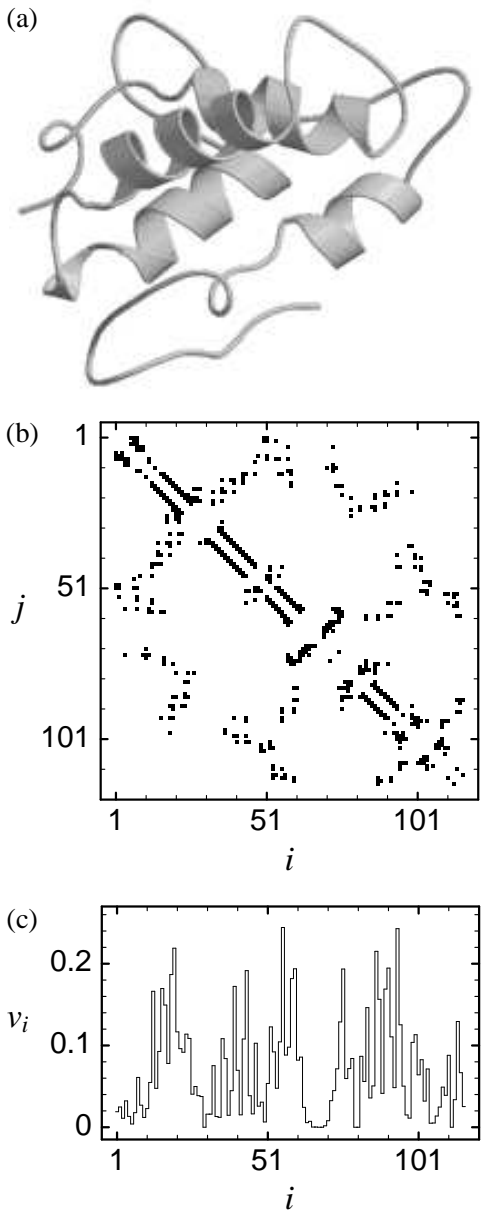}}
\caption{The protein Amylase/Serine protease inhibitor with PDB id.\
\texttt{1bea}. Shown are: (a)~The three-dimensional structure (drawn using
MolScript and Raster3D), (b)~the CM $\mathbf{C}$ (black means $C_{ij} = 1$ and
white means $C_{ij} = 0$), and (c)~the PE $\mathbf{v}$. The reduction from (a)
to (b) is done by a distance threshold, and the one from (b) to (c) is by a
diagonalization. The reconstruction from (c) to (b) is described in this
Letter, whereas the one from (b) to (a) is described in
Ref.~\protect\cite{vendruscolo97}.}
\label{figure:1bea}
\end{figure}

The PE of the CM has already been used as an indicator of protein topology, in
particular as a mean of identifying structural domains \cite{holm94} and
clusters of amino acids with special structural significance
\cite{kannan99,kannan00}. Here, we present an exact algorithm to reconstruct a
CM from the knowledge of its PE (cf.\ Fig.~\ref{figure:1bea}). This step is
analogous to that of reconstructing the three-dimensional protein structure
based on the CM of the structure \cite{vendruscolo97}. For the proteins that we
studied, the PE is sufficient to reconstruct uniquely a CM. This means that the
information about all other eigenvectors and eigenvalues of a CM is contained
in the PE, and therefore it is equivalent to represent a protein structure
either by the CM or the PE of the CM. This result is likely due to the binary
nature of the entries of the CM and to the fact that the topology of the
protein chain imposes significant constraints on the non-zero entries
\cite{vendruscolo99}.

\noindent\textit{Contact maps and their principal eigenvectors.} -- The contact
map $\mathbf{C}$ of a protein structure is a binary symmetric matrix, with
elements $C_{ij} = 1$ if amino acids at positions $i$ and $j$ are in contact,
and $0$ otherwise \cite{lifson79,vendruscolo97}. Two residues are defined to be
in contact if at least one pair of heavy atoms, one belonging to each amino
acid, are less than $4.5 \, \mbox{\AA}$ apart. Other contact definitions exist
in the literature, for instance based on a distance threshold on the
$\mathrm{C}_{\alpha}$ atoms \cite{vendruscolo97}, but the algorithm presented
below does not depend on the detailed contact condition. Additionally, only
residues separated by at least three positions along the sequence are
considered in contact, so that $C_{ij} = 0$ if $|i-j| < 3$. In such a way,
trivial short range contacts are not taken into account.

In what follows, $\lambda$ will denote the principal (i.e., largest) eigenvalue
of $\mathbf{C}$ \cite{england03}, and $\mathbf{v}$ the corresponding PE. Since
$\mathbf{C}$ is a real symmetric matrix, its eigenvalues are real. The
principal eigenvalue $\lambda$ has a value between the average number of
contacts per residue, $\big< \sum_j C_{ij} \big>_i$, and the maximal number of
contacts of any given residue, $\max_i \big( \sum_j C_{ij} \big)$
\cite{bollobas98}. The non-zero components of $\mathbf{v}$ have all the same
sign, which we choose to be positive.

The PE maximizes the quadratic form $\sum_{ij} C_{ij} \, v_i \, v_j$ with the
constraint $\sum_i v_i^2 = 1$. In this sense, $v_i$ can be interpreted as the
effective connectivity of position $i$, since positions with large $v_i$ are in
contact with as many as possible positions $j$ with large $v_j$. As we discuss
below, one can show that if the CM represents a single connected graph, all the
structural information is actually contained in its PE. For proteins consisting
of several distinct domains, the CM becomes a block matrix, and the PE contains
information only on the largest block (domain).

\noindent\textit{Reconstructing a contact map from its principal eigenvector.}
-- The reconstruction algorithm is based on imposing that the matrix
$\mathbf{C}$, applied to $\mathbf{v}$, fulfills the secular equation
$\mathbf{C} \mathbf{v} = \lambda \mathbf{v}$. At first glance, this equation
does not seem to be sufficient to determine the CM, as there are infinitely
many matrices $\mathbf{M} \not= \mathbf{C}$ that fulfill $\mathbf{M} \mathbf{v}
= \lambda \mathbf{v}$. However, additional constraints can be imposed using the
fact that the elements of $\mathbf{C}$ assume only the values $0$ and $1$, and
that consequently $\lambda$ as well as all non-zero components of $\mathbf{v}$
are positive. The central idea is to exploit these additional constraints, as
they allow to apply a `greedy' scheme to search for possible solutions. We
first discuss the algorithm in the hypothesis that all components of the
$\mathbf{v}$ and the corresponding principal eigenvalue $\lambda$ are known. We
will show later that, for CMs of protein folds, it is possible to deduce
$\lambda$ from the components of the PE, so that only this quantity has to be
known. The algorithm proceeds according to the following steps:

\noindent (i)~Elements $C_{ij}$ with $|i-j| < 3$ are set to $0$. The remaining
elements of $\mathbf{C}$ are marked as `unknown.'

\noindent (ii)~For all positions $i$ for which the PE vanishes, i.e.\ $v_i =
0$, all `unknown' entries in the $i$-th line and the $i$-th column of
$\mathbf{C}$ are set to $0$, i.e.\ $C_{ij} = C_{ji} = 0$ for all $j$
(concerning `unknown' elements $C_{ij}$ for which both $v_i$ and $v_j$ vanish
see below).

\noindent (iii)~The non-zero components of $\mathbf{v}$ are sorted and treated
recursively in increasing order, starting with the smallest value. Let $i$ be
the position presently examined, with the aim to determine the $i$-th line and
$i$-th column of $\mathbf{C}$. Some elements $C_{ij} = C_{ji}$ with $j \in
\mathcal{J} = \{ j_1, j_2, \ldots \}$ are already known, whereas some other
elements $C_{ik} = C_{ki}$ with $k \in \mathcal{K} = \{ k_1, k_2, \ldots \}$
are still `unknown.' To evaluate the latter, we calculate the sum $\sum_{j \in
\mathcal{J}} C_{ij} v_j$ of the known elements $\mathcal{J}$, and consider
three different possibilities:

\noindent (iii/1)~The sum $\sum_{j \in \mathcal{J}} C_{ij} v_j$ is equal to
$\lambda v_i$ up to the chosen precision $\epsilon$, i.e.\ $\big| \sum_{j \in
\mathcal{J}} C_{ij} v_j - \lambda v_i \big| < \epsilon$. Then one solution for
the $i$-th line and the $i$-th column of $\mathbf{C}$ has been found, since the
$i$-th line of the secular equation can be fulfilled by setting to $0$ all
`unknown' elements in the $i$-th line and the $i$-th column of $\mathbf{C}$,
$C_{ik} = C_{ki} = 0$ for $k \in \mathcal{K}$, and the recursion returns with
success.

\noindent (iii/2)~Otherwise, if either $\sum_{j \in \mathcal{J}} C_{ij} v_j >
\lambda v_i$, meaning numerically that $\sum_{j \in \mathcal{J}} C_{ij} v_j -
\lambda v_i \ge \epsilon$, or if the set of `unknown' entries $\mathcal{K}$ is
empty, the $i$-th line of the secular equation cannot be fulfilled with the
present set $\mathcal{J}$, the recursion is in a dead end and returns with
failure.

\noindent (iii/3)~Finally, in the remaining cases, the `unknown' elements in
the non-empty set $\mathcal{K}$ have to be further processed. To eliminate
candidates which lead to a dead end, the set of elements $\mathcal{K}$ is
sorted by the values $v_k$ in decreasing order. Starting with the element $k
\in \mathcal{K}$ which has the largest value $v_k$, the sum $\sum_{j \in
\mathcal{J}} C_{ij} v_j + v_k$ is calculated. There are two possibilities:
(a)~If $\sum_{j \in \mathcal{J}} C_{ij} v_j + v_k > \lambda v_i$, meaning
numerically that $\sum_{j \in \mathcal{J}} C_{ij} v_j + v_k - \lambda v_i \ge
\epsilon$, then it follows that $C_{ik} = C_{ki} = 0$, since all components of
$\mathbf{v}$ are positive. (b)~If $\sum_{j \in \mathcal{J}} C_{ij} v_j + v_k
\le \lambda v_i$, meaning numerically that $\sum_{j \in \mathcal{J}} C_{ij} v_j
+ v_k - \lambda v_i < \epsilon$, then $C_{ik} = C_{ki}$ is allowed to assume
both values $0$ and $1$, and the search branches. In both cases~(a) and (b),
the algorithm continues recursively until the set $\mathcal{K}' = \mathcal{K}
\setminus k$ of `unknown' entries is empty, and either a failure or a success
is reported.

After examining the $i$-th line in step~(iii), the algorithm has found all
possible binary matrices fulfilling the $i$-th line of the secular equation, as
well as the lines previously treated. For each of these partial solutions, the
reconstruction continues with the line corresponding to the next largest
component of $\mathbf{v}$. Thereby, a tree is constructed which grows by
another generation $t$ every time a new line of the secular equation is
evaluated. The leaves of the tree at generation $t$ are the partial solutions
that fulfill the so far treated $t$ lines of the secular equation. Some of the
leaves of the tree `die' in the step from generation $t-1$ to $t$, if no
solution for the presently treated line of the secular equation is found that
fulfills the previously treated $t-1$ lines of the secular equation as well.
The search continues either until all lines of the secular equation have been
treated, or until all leaves have `died.' In the former case, at least one
complete solution is found, whereas in the latter case the reconstruction stops
without result. Due to the particular order of the search, the exponentially
large number of possible CMs is explored in a `greedy' way, limiting the search
to the smallest possible subset by discarding wrong parts of the tree already
in an early stage of the search. For most of the proteins we have studied (see
below), the number of leaves in any given step did not exceed the value of
$10^4$ for a properly chosen value of the threshold $\epsilon$.

Notice that the elements $C_{ij} = C_{ji}$, $|i-j| > 2$, for which both $v_i$
and $v_j$ vanish are not determined by the PE. These positions belong to a set
which does not interact with the set of positions contributing to the PE,
either as a completely independent structural domain, or as an isolated residue
without any contact. In step~(ii), we set such values of $C_{ij} = C_{ji}$ to
$0$ for convenience. Their actual value can be determined at a later stage,
when the CM obtained through our algorithm is submitted to a procedure such as
the one of Ref.~\cite{vendruscolo97}, in order to determine the
three-dimensional structure. This second step yields the three-dimensional
structure, satisfying physical constraints, whose CM is most similar to the one
reconstructed from the PE.

If the principal eigenvalue $\lambda$ is not known, a simple way to guess its
value consists in taking the smallest non-zero PE component $v_i$, and use
(i)~all ratios $v_j/v_i$ with non-zero $v_j$ and $|i-j| > 2$, (ii)~all ratios
$(v_j+v_k)/v_i$ with non-zero $v_j$ and $v_k$, $k > j$, $|i-j| > 2$, and $|i-k|
> 2$, and (iii)~all ratios $(v_j+v_k+v_l)/v_i$ with non-zero $v_j$, $v_k$, and
$v_l$, $l > k > j$, $|i-j| > 2$, $|i-k| > 2$, and $|i-l| > 2$ as candidate
values. The first choice corresponds to assuming that position $i$ has only a
single contact, the second corresponds to assuming that position $i$ has two
contacts, whereas the third corresponds to assuming that position $i$ has three
contacts. Values larger than $10$ (for CMs on a $4.5\,\mbox{\AA}$ threshold on
the heavy atoms) are discarded. In this way, one can find a set of `guesses' of
the principal eigenvalue, which, for CMs of proteins, contains the correct
value. All wrong guesses for the principal eigenvalue get quickly discarded by
the algorithm, since the simulation runs into a dead end in these cases.

Another important issue concerns the choice of the numerical threshold
$\epsilon$. If the threshold is too small, it may be that no solution is found,
due to the numerical round-off error on the PE components. If it is too large,
the tree branches too often and the search becomes unmanageable. In our
calculations, values of order $\epsilon = 10^{-6}$ represented typically a good
compromise. Alternatively, as we have done in this study, the value of
$\epsilon$ can be automatically adjusted by letting it slowly increase,
starting with a small value, until a solution is found. Note that alternatively
to a threshold on the absolute error, it is possible to apply a threshold on
the relative error. When doing so, for instance the equality $\sum_{j \in
\mathcal{J}} C_{ij} v_j = \lambda v_i$ becomes $\big| \sum_{j \in \mathcal{J}}
C_{ij} v_j - \lambda v_i \big|/(\lambda v_i) < \epsilon$. The latter condition
is usually more appropriate if the entries $v_i$ are quite broadly distributed.

\noindent\textit{Results.} -- In order to assess the performance of the
algorithm, we set out to reconstruct the $221$ globular protein structures of
PDBselect25 of length $N \le 120$. We diagonalized the CMs of these $221$
proteins to obtain their PE. Then, we applied to each PE the reconstruction
scheme described above. In $205$ cases, the algorithm yielded a unique
solution, identical to the original native CM. Clearly, in these cases it is
possible to recover the three-dimensional protein structure with the same
accuracy achieved when starting with the native CM, for instance using the
scheme of Ref.~\cite{vendruscolo97}, with a typical root mean square
displacement (RMSD) of around $2\,\mbox{\AA}$. In the remaining $16$ cases, the
algorithm yielded a unique solution, yet the reconstructed CM differed from the
original one in one or several missing contacts, up to 8\% of the native ones,
which were undetermined since the corresponding pairs of components of the PE
are $0$. Nevertheless, the reconstructed CMs were very similar to the native
ones, so that it is still possible to recover the three-dimensional protein
structure with a high accuracy, as the RMSD was increased by 10\% or less with
respect to the structure obtained using the native CM. In all cases only a
\textit{single} CM was found and therefore the PE defines essentially in a
unique way a CM, and no false contact was contained in any of the obtained CMs.

It should be noted that the reconstruction of the CMs can differ considerably
in computational expenses, the three most difficult cases in the present set of
proteins being the CM of PDB id.\ \texttt{1gif\_A} ($N=115$), of PDB id.\
\texttt{1poa} ($N=118$), and of PDB id.\ \texttt{1bnk\_A} ($N=120$). For these
proteins, there are many almost identical non-zero components of the PE, so
that our strategy to efficiently eliminate dead ends was not very effective,
leading to an excessive branching of the search. However, even in these cases
the solution of the secular equation is unique and could be found with an
extensive search.

\noindent\textit{Discussion.} -- We have shown that the PE determines uniquely
the CM, apart from elements that correspond to pairs of positions with
vanishing PE components and are hence undetermined. This is a surprising
result, since it means that the remaining $N-1$ eigenvectors and eigenvalues of
the CM, where $N$ is of order $100$ or larger, are completely determined by the
PE. It can be understood by noting that the number of CMs is large but finite:
The total number of $N\times N$ symmetric binary matrices is $2^{N (N+1)/2}$,
whereas the total number of CMs representing protein structures increases with
$N$ as $\exp(a N)$, since the chain connectivity introduces correlations among
the contacts of neighboring residues \cite{vendruscolo99}. In contrast, the
number of PEs which are non-identical up to a given precision $\epsilon$
increases with decreasing $\epsilon$ as $\epsilon^{-N}$ without bound.
Therefore, for sufficiently small $\epsilon$, the number of these distinct PEs
is much larger than the number of CMs, so that the overwhelming majority of
them do not correspond to any CM, and we can expect that those which do
correspond to a CM correspond to a unique CM, as our results confirm. We note
that this is true, however, only for sufficiently accurate PEs. For noisy PEs,
the exact algorithm presented in this Letter is not suitable, and one has to
resort to a stochastic search, such as for example a Monte Carlo scheme. Such
scheme is considerably more complicated than the algorithm presented here,
mainly due to the existence of CMs which do not correspond to a physically
realizable structure but have a PE being almost identical to the one of the
target CM. Preliminary results suggest that the reconstruction of a noisy PE
might be possible when constraining the search to CMs corresponding to
protein-like structures with proper secondary structure and steric interaction.

The PE only contains information on the largest connected component of the
graph representing the protein structure. Nodes which are not connected with it
have vanishing PE components, and their mutual links are undetermined.
Therefore, our scheme is only able to determine the CM for the subset of
positions which have non-zero PE components. For single-domain globular
proteins, they represent the great majority of the positions in the chain, with
the only exceptions of small loops or single residues completely exposed to the
solvent. For non-globular proteins, whose CM is sparse and whose number of
contacts per residue is hence smaller than a threshold, the connected positions
are few and the method is able to yield only a portion of the CM. However, such
structures are problematic in any case, since their CM does not determine a
well defined three-dimensional structure, and they are not thermodynamically
stable in absence of other protein chains and other molecules with which they
interact (there are $27$ such protein structures in PDBselect25 of length $N
\le 120$ which we omitted in our study). More important difficult cases are
structures constituted of several almost independent domains. For such
structures, the PE components are very small outside the principal domain,
although they are not zero, since there is always a small number of
inter-domain contacts.

In conclusion, we have presented an exact algorithm that allows the CM of a
protein structure to be reconstructed from the sole knowledge of its PE. The
resulting CM can then be used to reconstruct the full three-dimensional
structure, for instance using the scheme of Ref.~\cite{vendruscolo97}. In this
sense, the three-dimensional structure of a protein fold can be reduced
(`compressed') into the PE of its CM, from which it can be recovered
(`decompressed') with no or minimal information loss. We have applied the
algorithm to the set of $221$ globular proteins of PDBselect25 of length $N \le
120$ and obtained in all cases a unique CM from the PE. In terms of structure
representation, our results show that a CM and its PE are equivalent, which
leads to a significant simplification in the representation of protein folds.
We anticipate that this result will create new possibilities in bioinformatics
applications, in particular those involving alignments of structure to
structure and of sequence to structure. Furthermore, our results have important
implications on our understanding of protein evolution. We have shown in fact
that the PE is correlated with the hydrophobicity profile of the amino acid
sequence attaining the fold, and even more correlated with the hydrophobicity
profile averaged over many sequences attaining the fold, so that protein
evolution can be understood as the motion of the hydrophobicity profile around
the PE of the fold \cite{bastolla04}. Finally, since the PE is related to the
contact vector \cite{kabakcioglu02}, which it seems to be possible to predict
from the protein sequence \cite{pollastri02}, a coarse prediction of the PE may
be possible as well and thus the PE may become an effective tool in protein
structure prediction approaches.

\bibliography{paper}

\end{document}